\documentclass[aps,prd,twocolumn,showpacs,amsmath,amssymb]{revtex4-1}
\usepackage{amsmath} \usepackage{graphicx} \usepackage{subfigure}
\usepackage{epstopdf} \usepackage{color} \usepackage{multirow}
\usepackage{setspace} \usepackage{overpic} \usepackage{amssymb} 
\usepackage{ulem}
\usepackage{float}
\usepackage[bookmarksnumbered, pdfstartview=FitH,colorlinks,urlcolor=blue, citecolor=blue,linkcolor=blue] {hyperref}
\usepackage{bm}
\usepackage{rotating}
\usepackage{xcolor}
\usepackage{makecell}
\usepackage{mathtext}
\usepackage{mathrsfs}
\usepackage[utf8]{inputenc}

\let\oldequation\equation
\let\oldendequation\endequation
\renewenvironment{equation}
 {\linenomathNonumbers\oldequation}
 {\oldendequation\endlinenomath}

\definecolor{boslv}{rgb}{0.0, 0.65, 0.58}
\definecolor{Munsell}{HTML}{00A877}

\newcommand{\psipp}{\psi(3686)}

\newcommand{\Br}{\mathcal{B}}

\newcommand{\chicJ}{\chi_{cJ}}

\newcommand{\GG}{\gamma\gamma}

\newcommand{\psp}{\psi(3686)}
\newcommand{\jpsi}{J/\psi}

\newcommand{\bfg}{\begin{figure}}
\newcommand{\efg}{\end{figure}}
\newcommand{\bitm}{\begin{itemize}}
\newcommand{\eitm}{\end{itemize}}
\newcommand{\bnum}{\begin{enumerate}}
\newcommand{\enum}{\end{enumerate}}
\newcommand{\btbl}{\begin{table}}
\newcommand{\etbl}{\end{table}}
\newcommand{\btbu}{\begin{tabular}}
\newcommand{\etbu}{\end{tabular}}
\newcommand{\bcl}{\begin{center}}
\newcommand{\ecl}{\end{center}}

\newcommand{\beq}{\begin{equation}}
\newcommand{\eeq}{\end{equation}}
\newcommand{\beqr}{\begin{eqnarray}}
\newcommand{\eeqr}{\end{eqnarray}}

\begin{document}


\title{\boldmath Measurement of the Branching Fraction for the Decay $\chi_{cJ}\to p\bar{p}\eta\pi^{0}$}

\author{
\begin{small}
\begin{center}
M.~Ablikim$^{1}$, M.~N.~Achasov$^{4,c}$, P.~Adlarson$^{76}$, O.~Afedulidis$^{3}$, X.~C.~Ai$^{81}$, R.~Aliberti$^{35}$, A.~Amoroso$^{75A,75C}$, Q.~An$^{72,58,a}$, Y.~Bai$^{57}$, O.~Bakina$^{36}$
% ...
\\
\vspace{0.2cm}
(BESIII Collaboration)\\
\vspace{0.2cm} {\it
$^{1}$ Institute of High Energy Physics, Beijing 100049, People's Republic of China\\
% ...
$^{3}$ Bochum  Ruhr-University, D-44780 Bochum, Germany\\
$^{4}$ Budker Institute of Nuclear Physics SB RAS (BINP), Novosibirsk 630090, Russia\\
% ...
$^{35}$ Johannes Gutenberg University of Mainz, Johann-Joachim-Becher-Weg 45, D-55099 Mainz, Germany\\
$^{36}$ Joint Institute for Nuclear Research, 141980 Dubna, Moscow region, Russia\\
% ...
$^{57}$ Southeast University, Nanjing 211100, People's Republic of China\\
$^{58}$ State Key Laboratory of Particle Detection and Electronics, Beijing 100049, Hefei 230026, People's Republic of China\\
% ...
$^{72}$ University of Science and Technology of China, Hefei 230026, People's Republic of China\\
% ...
$^{75}$ University of Turin and INFN, (A)University of Turin, I-10125, Turin, Italy; (B)University of Eastern Piedmont, I-15121, Alessandria, Italy; (C)INFN, I-10125, Turin, Italy\\
$^{76}$ Uppsala University, Box 516, SE-75120 Uppsala, Sweden\\
% ...
$^{81}$ Zhengzhou University, Zhengzhou 450001, People's Republic of China\\
\vspace{0.2cm}
$^{a}$ Deceased\\
% ...
$^{c}$ Also at the Novosibirsk State University, Novosibirsk, 630090, Russia\\
% ...
}\end{center}
\vspace{0.4cm}
\end{small}
}



\begin{abstract}
Using $(2712.4\pm 14.3)\times10^6 \psi(3686)$ events collected by the BESIII detector operating at the BEPCII collider, we present the first observations of the decays $\chi_{cJ}(J=0,1,2)\to p\bar{p}\eta\pi^{0}$. Their decay branching fractions are determined to be ${\cal B}(\chi_{c0}\to p\bar{p}\eta\pi^{0})=({2.41 \pm 0.07 \pm 0.19}) \times 10^{-4}$,  ${\cal B}(\chi_{c1}\to p\bar{p}\eta\pi^{0})=({1.95 \pm 0.05 \pm 0.12}) \times 10^{-4}$, and ${\cal B}(\chi_{c2}\to p\bar{p}\eta\pi^{0})=({1.31 \pm 0.05 \pm 0.08}) \times 10^{-4}$, where the first uncertainties are statistical and the second systematic.
\end{abstract}


\maketitle

\section{Introduction}
In the quark model, the $\chi_{cJ}(J=0,1,2)$ mesons are identified as $n^{2S+1}L_J=$ $1^3P_J$ charmonium states. Due to the principle of parity conservation, their direct production in $e^+e^-$ collisions is suppressed and occurs only through a two-photon exchange process, which results in their
low production rates in $e^+e^-$  annihilation.
 By contrast, the radiative decays of $\psipp$ into $\chicJ$ have branching fractions (BFs) of about $9\%$~\cite{pdg2022}, and could provide a sizable yield of 
  $\chicJ$ states. The BESIII Collaboration has collected about 2.7 billion $\psipp$ events since 2008~\cite{psip_num_0912},  thereby providing an excellent opportunity to investigate the $\chicJ$ properties with the large amounts of the $\chicJ$ mesons produced in the  $\psipp$ radiative decays. So far, the sum of all measured BFs of the $\chicJ$ decays are each still far less than 1. Intensive studies of their multi-body decays are lacking relative to the few-body decays, particularly for those multi-body decays that contains baryon pair. A search for new decay modes 
is useful in understanding the properties of $\chicJ$. Furthermore,  the $\chicJ$ mesons have different quantum numbers from $\jpsi$ and $\psi(3686)$, so the decays may contain additional information on the potential intermediate baryon and meson states, and also 
the mass threshold behavior of baryon pair in the multi-body system compared to that of the $\jpsi$ and $\psi(3686)$ decays.

In this paper, we investigate the multi-body decays $\chicJ\to p\bar{p}\eta\pi^{0}$ via the transition process $\psipp\to\gamma \chicJ$ with $(2712.4\pm14.3)\times10^6 ~\psipp$ events~\cite{psip_num_0912}. Additionally, the potential $p\bar{p}$ threshold enhancement, and the possible intermediate states are searched in the $\eta\pi^0$, $p\eta(\bar{p}\eta)$, and $p\pi^{0}(\bar{p}\pi^{0})$ mass spectrum.

\section{BESIII DETECTOR AND MONTE CARLO SIMULATION}
\label{sec:BES}

The BESIII detector~\cite{Ablikim:2009aa} records $e^+ e^-$ collisions provided by the BEPCII collider~\cite{CXYu_bes3}.~The cylindrical core of the BESIII detector covers 93\% of the full solid angle and consists of a helium-based multilayer drift chamber (MDC), a  time-of-flight system (TOF), and a CsI(Tl) electromagnetic calorimeter (EMC), which are all enclosed in a superconducting solenoidal magnet providing a 1.0~T magnetic field. The  solenoid is supported
by an octagonal flux-return yoke with resistive
plate muon identification modules interleaved with steel. The charged-particle momentum resolution at 1~GeV/$c$ is 0.5\%, and the  d$E/$d$x$ resolution is 6\% for the electrons from Bhabha scattering. The EMC measures photon energies with a resolution of 2.5\% (5\%) at 1~GeV in the barrel (end-cap) region. The time resolution in the plastic scintillator TOF barrel region is 68~ps, while that in the end cap region was 110~ps. The end-cap TOF system was upgraded in 2015 using multi-gap resistive plate chamber technology, providing a time resolution of 60~ps, which benefits $\sim83\%$ of the data used in this analysis~\cite{tof_a,tof_b,tof_c}.

Monte Carlo (MC) simulated data samples produced with a {\sc geant4}-based~\cite{geant4} software package, which includes the geometric description~\cite{detvis} of the BESIII detector and the detector response, are used to determine detection efficiencies and to estimate backgrounds.~The simulation models the beam energy spread and initial-state radiation in the $e^+e^-$ annihilation with the generator {\sc kkmc}~\cite{kkmc_a,kkmc_b}.~The inclusive MC sample includes the production of the $\psi(3686)$ resonance, the ISR production of the $J/\psi$, and the continuum processes incorporated in {\sc kkmc}.~Particle decays are generated by {\sc evtgen}~\cite{evtgen_a,evtgen_b} for the known decay modes with BFs taken from the Particle Data Group (PDG)~\cite{pdg2022} and by {\sc lundcharm}~\cite{lundcharm_a,lundcharm_b} for the unknown ones. Final-state radiation from charged final-state particles is included using the {\sc photos} package~\cite{photos}. 

The signal is simulated with the E1 transition $\psipp\to\gamma \chicJ$, where the polar angle $\theta$ of the radiative photon in the $e^+e^-$ center-of-mass frame is distributed according to $(1+\lambda \rm cos^2\theta)$. For $J$ = 0, 1, and 2, $\lambda$ is set to 1, $-\frac{1}{3}$, and $\frac{1}{13}$, respectively~\cite{ref1E1,ref2E1}. The decays $\chicJ\to p\bar{p}\eta\pi^0$, $\eta\to\gamma\gamma$ and $\pi^0\to\gamma\gamma$ are generated with the events evenly distributed in the phase space.

\section{EVENT SELECTION}
\label{sec:selection}

In this analysis, both $\eta$ and $\pi^{0}$ candidates are reconstructed by $\GG$ final states. Therefore, the total final state is $5\gamma p\bar{p}$.

Charged tracks detected in the MDC are required to be within a polar angle ($\theta$) range of $|\rm{cos\theta}|<0.93$, where $\theta$ is defined with respect to the $z$-axis,
which is the symmetry axis of the MDC. The distance of closest approach to the interaction point (IP) 
must be less than 10\,cm
along the $z$-axis, $|V_{z}|$,  
and less than 1\,cm
in the transverse plane, $|V_{xy}|$.  The event candidates must have two 
good charged tracks with opposite charges.
Particle identification~(PID) for charged tracks combines measurements of the energy deposited in the MDC~(d$E$/d$x$) and the flight time in the TOF to form likelihoods $\mathcal{L}(h)~(h=p,K,\pi)$ for each hadron $h$ hypothesis.
The tracks are identified as protons when the proton hypothesis has the greatest likelihood ($\mathcal{L}(p)>\mathcal{L}(K)$ and $\mathcal{L}(p)>\mathcal{L}(\pi)$). 
Finally, the events containing one proton and one anti-proton are retained.

Photon candidates are identified using isolated showers in the EMC. The deposited energy of each shower must be greater than 25 MeV in the barrel region ($|\cos\theta|<0.80$) and more than 50 MeV in the end-cap regions  ($0.86<|\cos\theta|<0.92$). To suppress electronic noise and showers unrelated to the event, the difference between the EMC time and the event start time is required to be within [0,700] ns. The number of photon candidates is required  to be at least five in an event.

In order to suppress backgrounds and improve the mass resolution, a six-constraint (6C) kinematic fit is performed with the $\psp\to\gamma p\bar{p}\eta\pi^{0}$ hypothesis by constraining the total four-momentum of the final state particles to that of the colliding beams, and the extra 2C is used to constraint the invariant masses of the photon pairs to the $\eta$ and $\pi^{0}$ nominal masses~\cite{pdg2022}, respectively. If the 6C kinematic fit converges for more than one combination of the photons, the one with the least $\chi^2_{\rm 6C}$ is chosen. Furthermore, the selection $\chi^{2}_{\rm 6C}<20$ is applied, by optimizing a figure of merit defined as $S/\sqrt {S+B}$, where $S$ denotes the number of signal events obtained from the MC simulation that scaled by the following normalized formula,
{\footnotesize
${N}^{\text{normalized}} = \epsilon_{\chi_{c0}} \cdot N_{\psi(3686)}^{\text{Data}} \cdot {\cal B}\left(\psi(3686) \to \gamma \chi_{c0} \right) \cdot  {\cal B}\left( \eta \to \gamma \gamma \right) \cdot  {\cal B}\left( \pi^{0} \to \gamma \gamma \right) \cdot  {\cal B}\left (\chi_{c0} \to p\bar{p} \eta\pi^{0} \right)$},
where the $\chi_{c0}\to\bar p p\eta\pi^0$ is the measured branching fraction in this analysis, while $B$ is the number of background events obtained from the inclusive MC sample. 
Moreover, in order to suppress the backgrounds with four or six photons, a 4C kinematic fit is performed for both signal and background channels by looping all photon combinations according to the different numbers of photons hypothesis. The one with the least $\chi^2$ is kept for each channel. Then, we require
\begin{align}
    \chi^{2}_{\rm4C}(5\gamma p\bar{p})<\chi^{2}_{\rm4C}(4\gamma p\bar{p}),
\end{align}
\begin{align}
\chi^{2}_{\rm4C}(5\gamma p\bar{p})<\chi^{2}_{\rm4C}(6\gamma p\bar{p})
\end{align}
To suppress backgrounds of $\chi_{cJ}\to (\gamma\gamma) p\bar{p}\eta\eta(\pi^{0}\pi^{0})$, we also requre
\begin{align}
\chi^{2}_{\rm6C}(\gamma p\bar{p}\eta\pi^{0} )<\chi^{2}_{\rm6C}(\gamma p\bar{p}\eta\eta ),
\end{align}
\begin{align}
\chi^{2}_{\rm6C}(\gamma p\bar{p}\eta\pi^{0})<\chi^{2}_{\rm6C}(p\bar{p}\eta\eta ),
\end{align}
\begin{align}
\chi^{2}_{\rm6C}(\gamma p\bar{p}\eta\pi^{0})<\chi^{2}_{\rm6C}(\gamma p\bar{p}\pi^{0}\pi^{0} ),
\end{align}
\begin{align}
\chi^{2}_{\rm6C}(\gamma p\bar{p}\eta\pi^{0} )<\chi^{2}_{\rm6C}(p\bar{p}\pi^{0}\pi^{0} ),
\end{align}
\begin{align}
\chi^{2}_{\rm6C}(\gamma p\bar{p}\eta\pi^{0})<\chi^{2}_{\rm6C}(\gamma\gamma p\bar{p}\pi^{0}\pi^{0})
\end{align}.

After the aforementioned selection criteria, it is found that a large background component comes from the $J/\psi$-related channel. They are suppressed by requiring the $\eta$ recoil mass (RM) and the invariant mass of $p\bar{p}\eta$ to be outside the $J/\psi$ mass windows. The optimized mass windows are determined to be $\left|RM_{\eta} - m_{J/\psi}\right|>7.2{\rm ~MeV}/c^2$, and  $\left|M_{p\bar{p}\eta}
- m_{J/\psi}\right|>21.3{\rm ~MeV}/c^2$, where $m_{J/\psi}$ is the $\jpsi$ nominal mass~\cite{pdg2022}.

The remaining background from the process $\chi_{cJ}\to p\bar{p}\pi^0\pi^0$ is suppressed by requiring the $M_{\gamma\gamma}$ to be outside the $\pi^0$ mass window.~ We checked the invariant mass spectrum of all  $\gamma\gamma$ combinations for the five candidate photons in ranking with their energies, and found four of them form the $\pi^0$ signals. The requirements,  $\left|M_{\gamma_{1}^{\pi^{0}}\gamma_{1}^{\eta }} - m_{\pi^0}\right|>14.1~{\rm MeV}/c^2$, $\left|M_{\gamma_{1}^{\pi^{0}}\gamma_{2}^{\eta }} -m_{\pi^0}\right|>15~{\rm MeV}/c^2$ ,  $\left|M_{\gamma_{1}^{\eta }\gamma^{E}} -m_{\pi^0}\right|>21.3~{\rm MeV}/c^2$, and $\left|M_{\gamma_{2}^{\eta }\gamma^{E}} -m_{\pi^0}\right|>~{\rm 19.5 ~MeV}/c^2$, are performed to remove the $p\bar{p}\pi^0\pi^0$  backgrounds, where the $\gamma^{E}$ represents radiative photon. 

The other remaining potential backgrounds are investigated with the inclusive $\psi(3686)$ MC samples, using the event-type analysis tool TopoAna~\cite{zhouxy_topoAna}.  It is found that there are several main background channels with the contributions larger than 1$\%$ as in Table~\ref{normalized_value}. The efficiencies are obtained using the large exclusive MC samples, and the normalized numbers of background events are calculated by the following formula,
{\footnotesize
${N}^{\text{normalized}} = \epsilon_{bkg} \cdot N_{\psi(3686)}^{\text{Data}} \cdot \text{Br}(bkg)$}, where the $\text{Br}(bkg)$ are the BFs of these different background processes in PDG~\cite{pdg2022}.  The contributions of these backgrounds will be taken into account in the fit to extract the signal yields later.

\begin{table}[htp]
\begin{center}
\caption{The estimated numbers of the peaking background events, $N_2$, and their efficiencies.}
\label{normalized_value}
\begin{tabular}{c c c}
\hline
Mode                                                    &  Efficiency~(\%) & $N_2$  \\
\hline
$\chi_{c1}\to\gamma J/\psi, J/\psi \to \eta p\bar{p} $  &  0.08         &   ${53.9\pm7.0}$                          \\
$\chi_{c2}\to\gamma J/\psi, J/\psi \to \eta p\bar{p}$   &  0.05          &  ${20.1\pm2.7}$                   \\
$\chi_{c0}\to\pi^0\pi^0 p\bar{p}$                       &  0.01         &  
${36\pm11}$                       \\
$\chi_{c2}\to\pi^0\pi^0 p\bar{p}$                       &  0.01         &  ${14.1\pm4.7}$                            \\
$\chi_{c1}\to\gamma J/\psi, J/\psi\to\omega p\bar{p}$           &  0.70          &  ${51\pm11}$             \\
$\chi_{c2}\to\gamma J/\psi, J/\psi \to\omega p\bar{p}$           &  0.49         &   ${19.4\pm4.2}$             \\
\hline
\end{tabular}
\end{center}
\end{table}

The continuum background is investigated using the data samples collected at center-of-mass energy of 3.65 GeV with an integrated luminosity of 454 $\mathrm{pb^{-1}}$~\cite{cont}.  There are five events passing the selection criteria, so the contribution from the continuum process is neglected.

Figure~\ref{fit_result} shows the $M(p\bar{p}\eta\pi^0)$ distribution with the survived events in data, and clear $\chicJ$ signals are seen. The $\chi_{c0}$, $\chi_{c1}$, and $\chi_{c2}$ signal regions are defined as $[3.38, 3.45]$, $[3.49, 3.53]$, and $[3.54, 3.58]$ GeV/$c^{2}$, respectively.
\section{\label{Sec:BR_determined}Signal yields}
\begin{figure*}[htbp]
		\begin{minipage}[t]{0.6\linewidth}
		\includegraphics[width=1\textwidth]{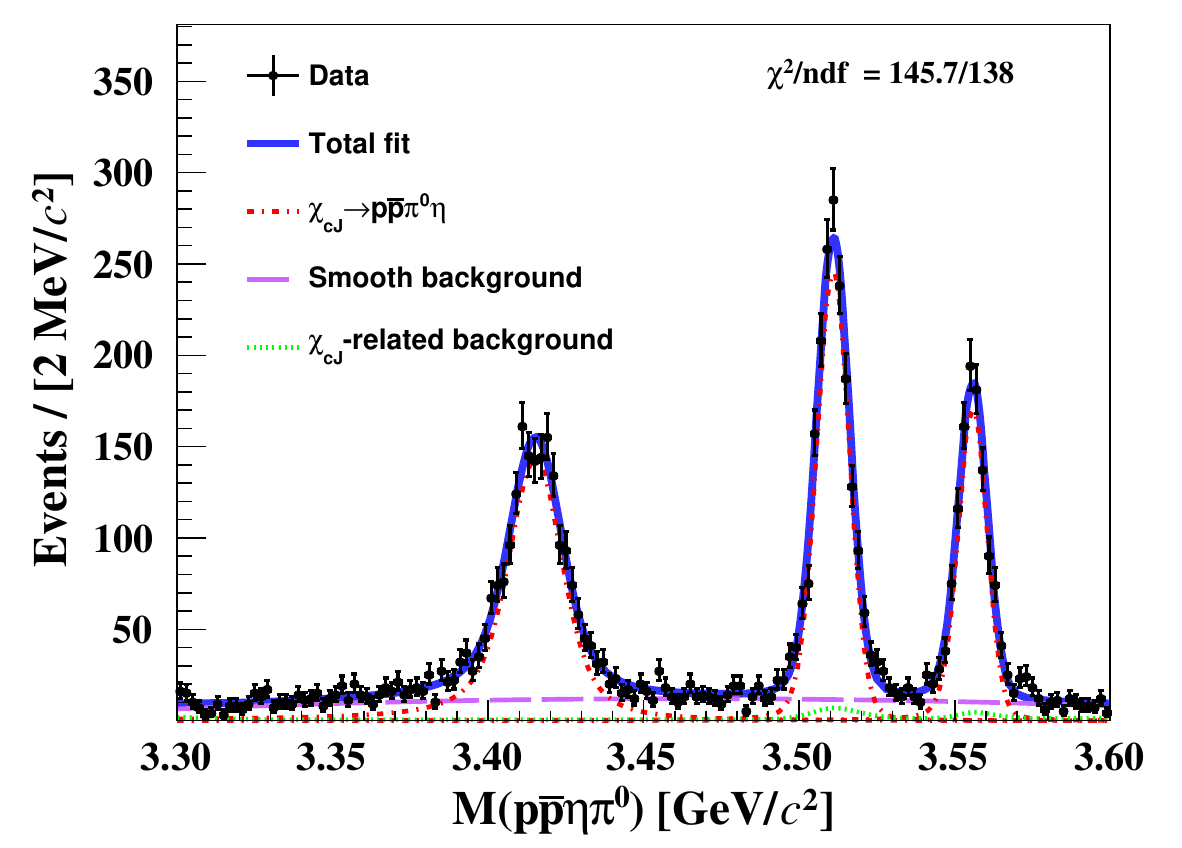}
		\end{minipage}
		
\caption{ The fit to the M$(p\bar{p}\eta\pi^{0})$ distribution from data, where the points with error bars denote the data, the blue line denote the total fit, the red dashed line denote the signal shape, the purple line denote the smooth background, and the green dashed line denote the $\chi_{cJ}$-related background.}
		\label{fit_result}
		\end{figure*}

 To determine signal yields, an unbinned maximum likelihood fit is performaed on M$(p\bar{p}\eta\pi^{0})$ distribution with the accepted candidates in data. The $\chi_{cJ}$ signals described as 
\begin{align}
PDF_{\rm signal}=(E_{\gamma}^3\times BW(M)\times f_d(E_{\gamma}))\otimes G(\delta m,\sigma), 
\end{align}
 Here, the mean and width of the Gaussian function are free for all $\chi_{cJ}$ signals. And $E_{\gamma}=(m_{\psi(3686)}^{2}-m^2)/{2m_{\psi(3686)}}$ is the energy of the E1 transition photon, where $m_{\psi(3686)}$ is the $\psi(3686)$ nominal mass~\cite{pdg2022}. $BW(M)=1/({(M-m_{\chicJ})^2+0.25\Gamma^2_{\chicJ}})$ is the Breit-Wigner function, where $m_{\chicJ}$ and $\Gamma_{\chicJ}$ are the $\chicJ$ mass and width, respectively, and fixed to the values in the PDG~\cite{pdg2022}. The function $f_{d}(E_{\gamma})$ is used to damp the diverging tail caused by the $E_{\gamma}^3$ term, and is given by $(E_{0}^{2}/(E_{\gamma}E_{0}+(E_{\gamma}-E_{0})^{2}))$, as introduced by the KEDR experiment~\cite{damping}, where $E_{0}=(m_{\psi(3686)}^{2}-m_{\chi_{cJ}}^2)/{2m_{\psi(3686)}}$ denotes the peaking energy of the transition photon. 
The smooth background is described with a $2^{nd}$-order Chebychev polynomial. The $\chicJ$-related backgrounds are described by the MC-simulated shapes, and the number of background events are fixed to the values listed in Table~\ref{normalized_value}. 
Figure~\ref{fit_result} shows the fitting result.
The fit goodness is calculated to be $\chi^2/n.d.f.$ = 145.7/138. The obtained signal yields are summarized in Table~\ref{list_summary}.
 


\begin{table*}[htbp]
\begin{center}
\caption{ The signal yields $N_{\chi_{cJ}}^{\mathrm{obs}}$, detection efficiencies, and $\Br\left(\chi_{cJ} \rightarrow p\bar{p}\eta\pi^{0}\right)$, where the first uncertainties are statistical and the second systematic.}
\label{list_summary}
\begin{tabular}{c c c c}
\hline    
\multicolumn{2}{c}{$N_{\chi_{cJ}}^{\mathrm{obs}}$}& $\epsilon(\chi_{cJ}\to p\bar{p}\eta\pi^{0}) $~(\%)&$\Br(\chi_{cJ} \to p\bar{p}\eta\pi^{0})~$($\times 10^{-4}$) \\
\hline 
$\chi_{c0}$ & $1897.8\pm55.7$  &  $7.62\pm0.2$ & $2.41\pm0.07\pm0.19$ \\
$\chi_{c1}$ & $1694.6\pm46.7$  &  $8.44\pm0.2$ & $1.95\pm0.05\pm0.12$  \\
$\chi_{c2}$ & $1058.9\pm37.2$ &  $8.04\pm0.2$ & $1.31\pm0.05\pm0.08$ \\

\hline
\end{tabular}
\end{center}
\end{table*}

\begin{figure*}[htbp]
\centering
\begin{minipage}[t]{0.45\linewidth}
\includegraphics[width=\textwidth]{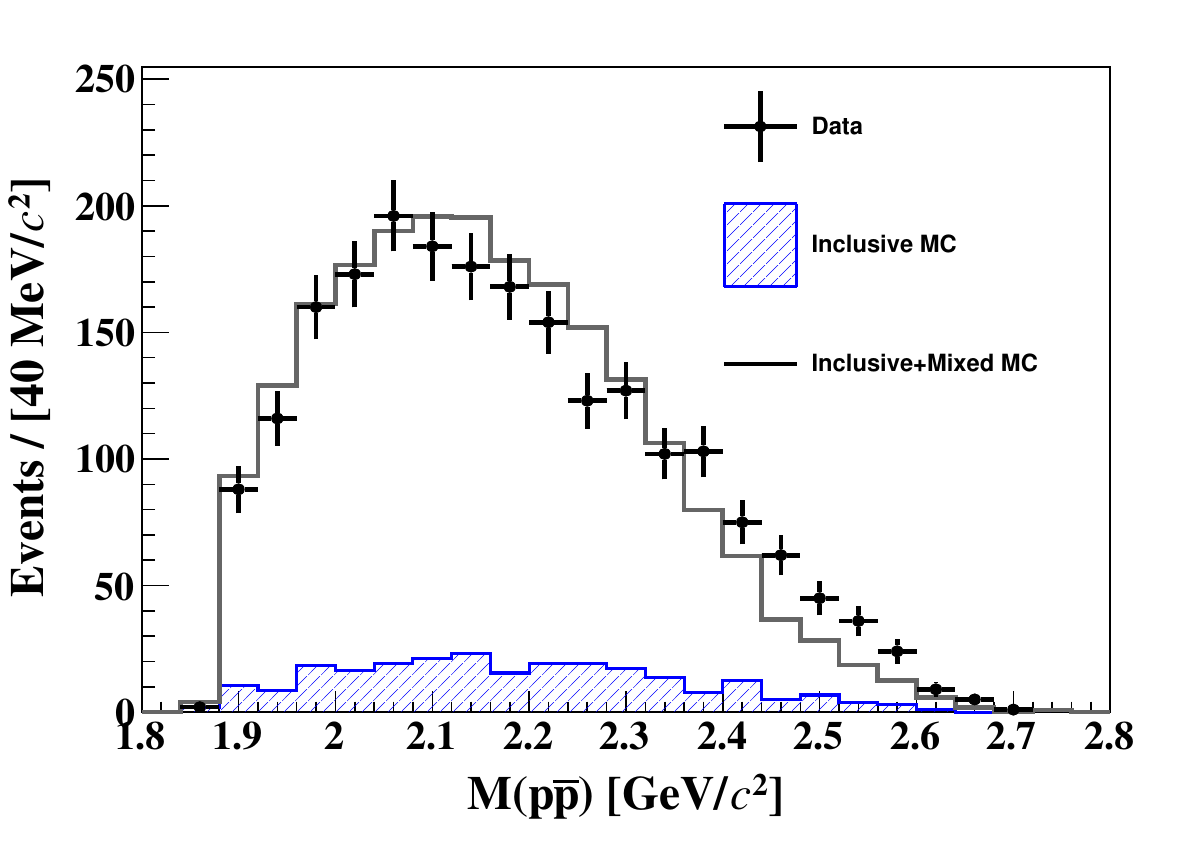}
\end{minipage}
\begin{minipage}[t]{0.45\linewidth}
\includegraphics[width=\textwidth]{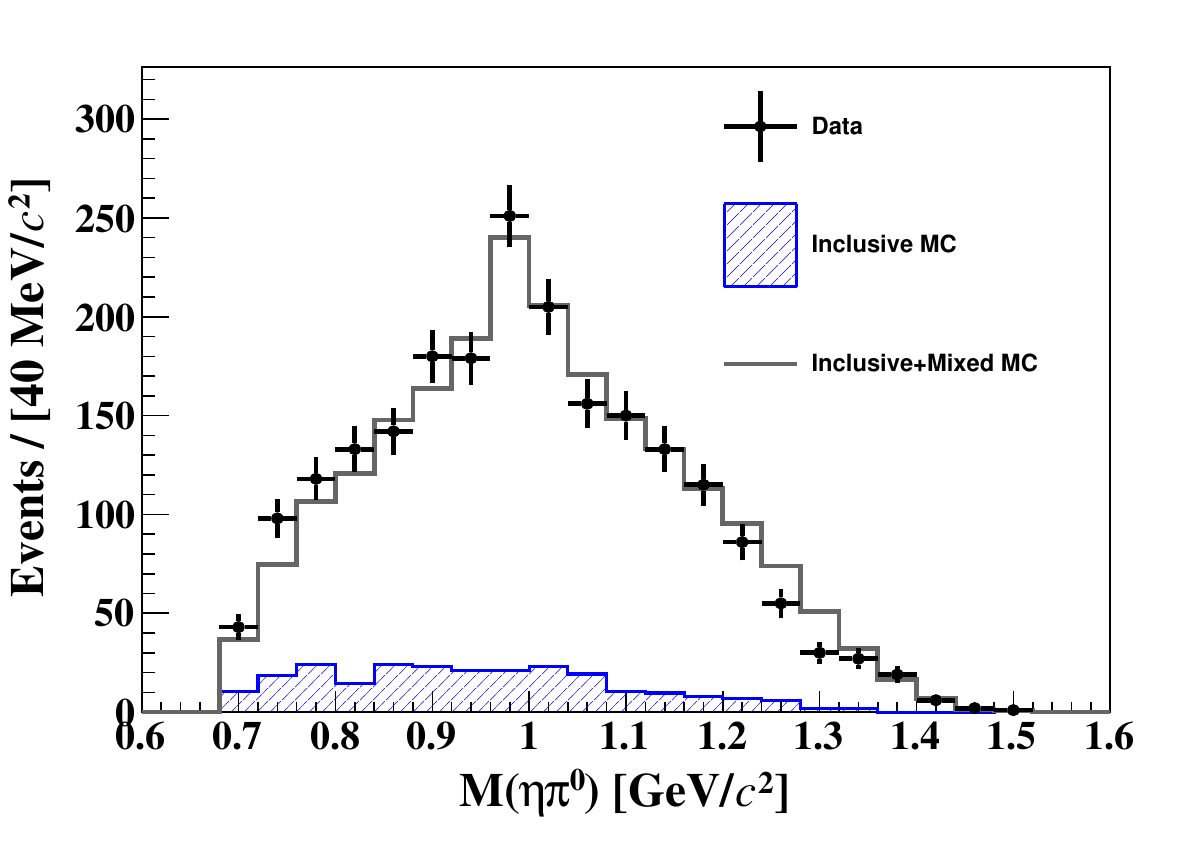}
\end{minipage}
\begin{minipage}[t]{0.45\linewidth}
\includegraphics[width=\textwidth]{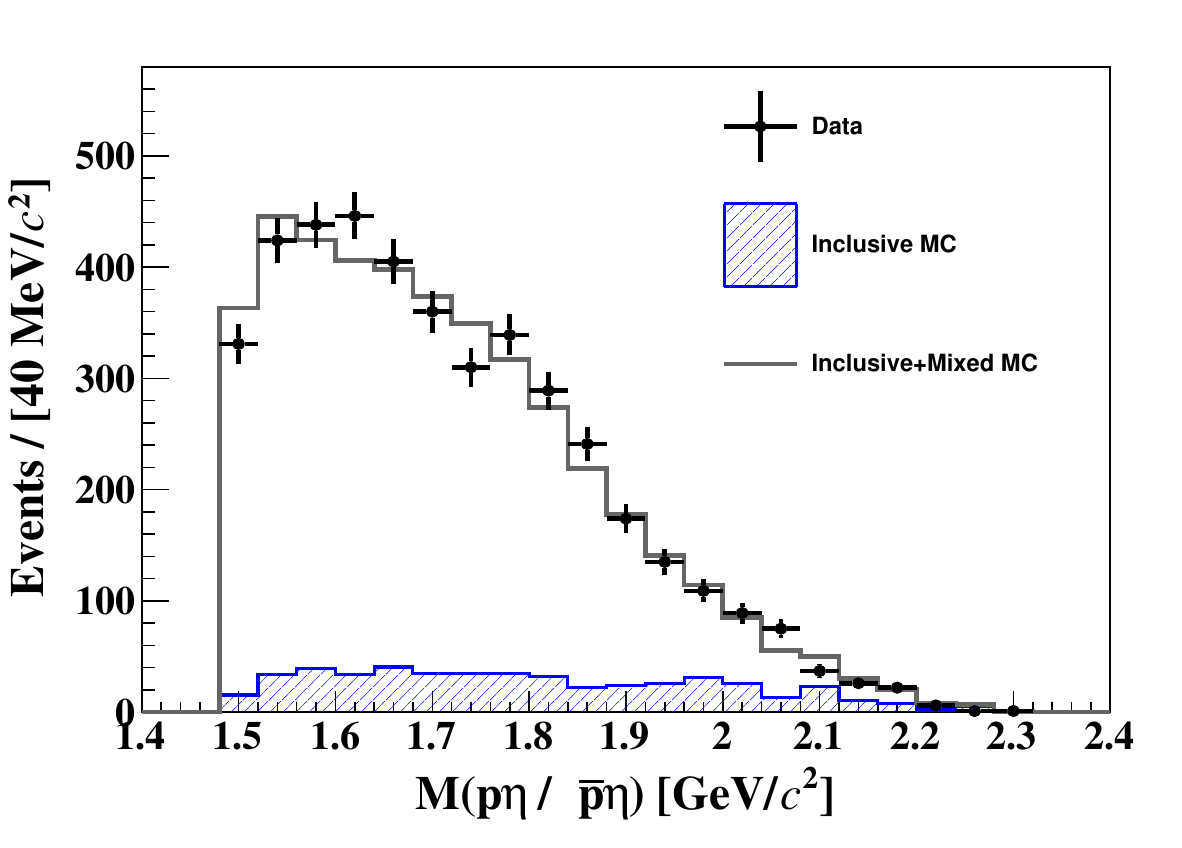}
\end{minipage}
\begin{minipage}[t]{0.45\linewidth}
\includegraphics[width=\textwidth]{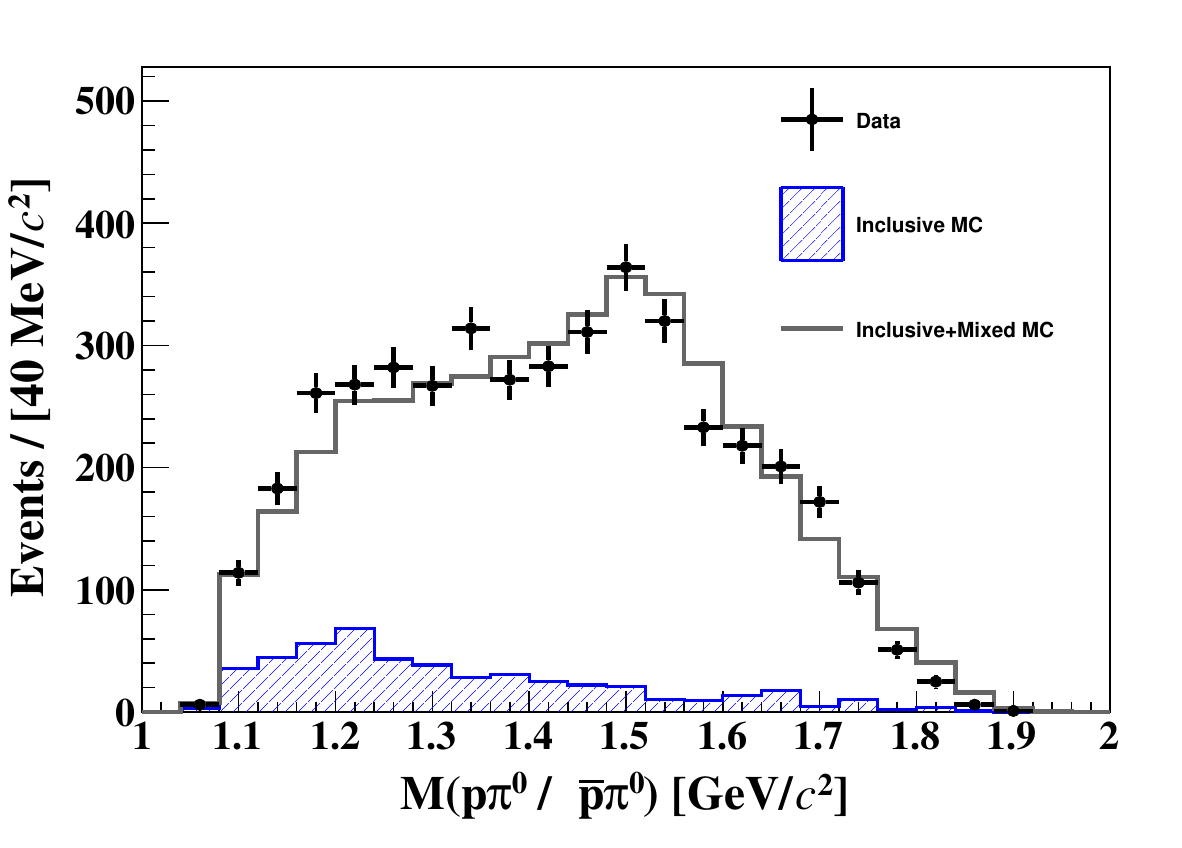}
\end{minipage}
\caption{Comparisons of the distributions of the invariant masses of two-body combinations of the accepted candidates for $\chi_{c0}\to p\bar{p}\eta\pi^0$. The points with error bars denote the data, the shaded histograms denote the background from inclusive MC sample, and the histogram denote the mixed MC sample plus the background contribution.}
\label{inter_chic0}
\end{figure*}

The potential intermediate states are investigated in the $p\bar{p}$, $\eta\pi^{0}$, $p\eta/\bar{p}\eta$ and $p\pi^{0}/\bar{p}\pi^{0}$ invariant mass spectra as shown in Figure~\ref{inter_chic0}. 
Here, we take $\chi_{c0}$ as an example. The hints of the scalar meson $a_0(980)$ and the excited baryon $N(1535)$ can be seen. But no significant $p\bar{p}$ threshold enhancement is observed. The numbers of $a_0(980)$ and $N(1535)$ are obtained via one or two-dimensional mass spectrum fits, respectively. The one-dimensional fit to M($\eta\pi^0$) is performed to determine the proportions of $a_{0}(980)$, while the two-dimensional fits to M($p\pi^{0}$) vs M($\bar{p}\eta$) and M($\bar{p}\pi^{0}$) vs M($p\eta$) are performed to determine the proportions of N(1535). Here, the proportions of $\chicJ\to p\bar{p}a_0(980)$ are approximately 10$\%$, 10$\%$ and 10$\%$  in the $\chicJ$ signal regions, the proportions of $\chicJ\to N(1535)\bar{p}\pi^0+c.c.$ are approximately 30$\%$, 30$\%$ and 30$\%$ and $\chicJ\to N(1535)\bar{p}\eta+c.c.$ are approximately 30$\%$, 40$\%$ and 30$\%$. These fractions are used to generated the mixed MC.  

The BFs of $\chi_{cJ}{(J=0,1,2)} \to p\bar{p}\eta\pi^{0}$ are calculated as
\begin{align}
    \Br({\chi _{cJ}} \to p\bar{p}\eta\pi^{0} ) =
\frac{N_{\chicJ}^{\rm obs}}{N_{\psi(3686)}\cdot\Br\cdot\varepsilon},
\end{align}
where $N_{\chi_{cJ}}^{\mathrm{obs}}$ is the signal yield of $\chi_{cJ}\to p\bar{p}\eta\pi^0$, $N_{\psi(3686)}$ is the number of $\psi(3686)$ events,  
$\Br=\Br(\psi(3686)\to\gamma\chicJ)\cdot\Br(\eta\to\gamma\gamma)\cdot\Br(\pi^0\to\gamma\gamma)$, and $\varepsilon$ denotes the detection efficiencies obtained from the mixed signal MC samples. The measured BFs are summarized in Table~\ref{list_summary}.

\section{Systematic Uncertainties}\label{sec:sysU}
The systematic uncertainties in the BF measurements originate from several sources, as summarized in Table~\ref{list_sys}. They are introduced below.

\begin{itemize}

\item[(I)]{{\bf Tracking:}}
The uncertainty of the $p\bar{p}$ tracking is estimated by the control sample $J/\psi \to p\bar{p}\pi^+\pi^-$, and is assigned as $1.0\%$ for each $p/\bar{p}$~\cite{Yuan:2015wga}.

\item[(II)]{\bf PID:} The uncertainty of the PID of $p\bar{p}$ is studied using the control sample $e^+e^-\to p\bar{p}\pi^0$, and is estimated to be $1.0\%$ for each $p/\bar{p}$~\cite{BESIII:2012urf}.

\item[(III)]{\bf Photon selection:}
The uncertainty of the photon selection is studied using the control sample $e^{+}e^{-} \to \gamma\mu^{+}\mu^{-}$~\cite{BAM662}, and is determined to be $0.5\%$ for each photon.

\item[(IV)]{\bf \boldmath$\eta$ and $\pi^{0}$ reconstruction:} The uncertainty from the $\eta$ and $\pi^{0}$ reconstruction is $1.0\%$ per $\eta$ or $\pi^{0}$, studied with the high purity control samples $J/\psi \to p\bar{p}\eta$ and $J/\psi \to \pi^{+}\pi^{-}\pi^{0}$~\cite{BESIII:2020}.

\item[(V)]{\bf Kinematic fit}: The systematic uncertainty of kinematic fit is estimated by comparing the efficiencies with and without applying the helix correction~\cite{helix}. The correction factors for protons are obtained by studying the control sample $\psipp \to p\bar{p}\pi^0$. The systematic uncertainties are estimated to be 1.9$\%$, 2.0$\%$, and 1.9$\%$ for $\chi_{cJ}(J=0,1,2) \to p\bar{p}\eta\pi^{0}$, respectively.

\item[(VI)]{\bf \boldmath Signal yield}
\item{\bf \boldmath Mass window}:
We set $\pi^0$ and $J/\psi$ mass windows to veto those related backgrounds as mentioned before.
To estimate the systematic uncertainties caused by these requirements, we examine the branching fractions by adjusting the relevant mass window. For each background veto, we vary the corresponding mass windows nine times with a step of 0.2 or 0.5 MeV/$c^2$. Here, 0.2 MeV/$c^2$ corresponds to the $J/\psi$ recoil mass window, while 0.5 MeV/$c^2$ corresponds to the $J/\psi$ mass window in $M(p\bar{p}\eta))$  and $\pi^0$ mass windows. For each case, the deviation between the alternative and nominal one is defined as $\zeta=\frac{\left|\Br_{\text {nominal }}-\Br_{\text {test }}\right|}{\sqrt{\left|\sigma_{\Br\text{,nominal }}^2-\sigma_{\Br\text {,test}}^2\right|}}$, where $\Br$ denotes branching fractions of $\chi_{cJ}\to p \bar{p}\eta\pi^0$ and  $\sigma$ denotes statistical uncertainty. If $\zeta$ is less than 2.0, the associated systematic uncertainty is negligible according to the Barlow test~\cite{barlow_test}. Otherwise, its relative difference is assigned as the systematic uncertainty. 

\item{\bf Fit method}:

The systematic uncertainties due to the fit range are examined by varying the nominal fit range from [3.29, 3.60] to [3.31, 3.60] GeV/$c^2$, with a step of 2 MeV/$c^2$, and the Barlow test is performed as above. The systematic uncertainties are $1.3\%$, $3.0\%$ and $2.7\%$ for $\chi_{cJ}(J=0,1,2)\to p\bar{p}\eta\pi^{0}$, respectively.

The uncertainty caused by the signal shape is estimated by replacing the damping factor with an alternative damping factor used by CLEO \cite{damping1}, $f_{d}({\it{E}}_{\gamma}) = {\rm {exp}}(-E^{2}_{\gamma}/8\beta^{2})$, is chosen to estimate the uncertainty from the damping function, where $\beta$ is a free parameter. The difference between the two damping functions is taken as the systematic uncertainty. They are $0.1\%$, $0.3\%$, and $1.2\%$ for $\chi_{cJ}(J=0,1,2) \to p\bar{p}\eta\pi^{0}$, respectively.

The systematic uncertainty caused by peaking background is estimated by varying the number of events by one standard deviation, which includes the statistical uncertainty of the MC simulated sample and the uncertainty related to the BF.~The largest differences relative to the nominal BF are assigned as the corresponding systematic uncertainty.

The uncertainty due to the non-peaking background shape is estimated by replacing the  $2^{\rm nd}$-order Chebychev polynomial with the $3^{\rm rd}$-order Chebychev one. The difference relative to the nominal result is taken as the corresponding systematic uncertainty.

\item[(VII)]{\bf  Quoted BFs}:
The uncertainties due to the quoted BFs of the intermediate states are cited from from the PDG~\cite{pdg2022}.

\item[(VIII)]{\bf MC statistics}:
The uncertainty due to the limited MC statistics is calculated by $\sqrt{\dfrac{(1-\varepsilon)\varepsilon}{N}}$, where $\varepsilon$ is the detection effciency and $N$ is the total number of MC events. These uncertainties are all $0.2\%$ for three decay final states.

\item[(IX)]{\bf  MC model}:
The systematic uncertainty due to intermediate states is estimated by comparing the nominal detection efficiency with that obtained with the phase space model, and the differences are determined to be 2.7$\%$, 1.1$\%$, and 0.5$\%$ for $\chi_{cJ}(J=0,1,2)\to p\bar{p}\eta\pi^{0}$, respectively, which are taken as the systematic uncertainties.

\item[(X)]{\bf Intermediate states}:
The systematic uncertainty due to fitting fractions of intermediate states is estimated by varying the fractions. The largest differences relative to the nominal BFs are assigned as the corresponding systematic uncertainty.

\item[(X)]{\bf \boldmath$N_{\psi(3686)}$}:
The uncertainty from the total number of $\psi(3686)$ events is determined to be 0.5\%~\cite{psip_num_0912}.

\end{itemize}
All the systematic uncertainties are summarized in Table~\ref{list_sys}, where the total uncertainties are obtained by adding all of them in quadrature, assuming all contributions are independent.
\begin {table}[H]
\begin{small}
\renewcommand\arraystretch{1.2}
{\caption {Relative systematic uncertainties in the measurements of the BFs of $\chi_{cJ}\to p\bar{p}\eta\pi^{0}$, where the `/' means negligible.}
\label{list_sys}}
\begin {tabular}{l c c c}

\hline
Source                    & $\chi_{c0}~(\%)$ & $\chi_{c1}~(\%)$ & $\chi_{c2}~(\%)$ \\
\hline
Tracking                  & 2.0         & 2.0         & 2.0         \\
PID                       & 2.0         & 2.0         & 2.0         \\
Photon                    & 2.5         & 2.5         & 2.5         \\
$\eta$ reconstruction     & 1.0         & 1.0         & 1.0         \\
$\pi^{0}$ reconstruction  & 1.0         & 1.0         & 1.0         \\
Kinematic fit             &1.9         &  2.0         & 1.9        \\
$\pi^0$  mass window       &  /        & /         &  /         \\
RM$(\eta)$ veto region       & 0.6         & /         &  /         \\
M$(p\bar{p}\eta)$ veto region       & 1.0         &0.5         & 0.4         \\
Fitting range           & 1.3         & 3.0         & 2.7         \\
Signal shape              & 0.1         & 0.3         & 1.2         \\
Non-peaking  background          & 3.5         & 0.9         & 0.2         \\
Peaking  background          & 3.4         & 1.4         & 0.3         \\
Quoted BFs                & 2.1         & 2.5         & 2.2         \\
MC statistics             &0.2           &0.2           &0.2        \\
MC model        & 2.7        & 1.1           &0.5   \\
Intermediate states & 0.8              &  0.4        &  0.1         \\
$N_{\psipp}$              & 0.5         & 0.5         & 0.5         \\
\hline
Total                     & 7.7        &6.3         & 5.8        \\
\hline

\end{tabular}
\end{small}

\end{table}

\section{Summary}
In summary, based on $(2712.4\pm 14.3)\times10^6\ \psipp$ events taken by the BESIII detector, the decays $\chi_{cJ}(J=0,1,2)\to p\bar{p}\eta\pi^{0}$ are observed for the first time. The corresponding BFs are determined to be $\Br(\chi_{c0} \to p\bar{p}\eta\pi^{0})$ $=({2.41 \pm 0.07 \pm 0.19}) \times 10^{-4}$,  $\Br(\chi_{c1} \to p\bar{p}\eta\pi^{0})$ $=({1.95 \pm 0.05 \pm 0.12}) \times 10^{-4}$, and $\Br(\chi_{c2} \to p\bar{p}\eta\pi^{0})$ $=({1.31 \pm 0.05 \pm 0.08}) \times 10^{-4}$, respectively. The observation for this new multi-body decay that contains baryon pair
will be useful in understanding the properties of $\chicJ$. 

\section{ACKNOWLEDGMENTS}
The BESIII Collaboration thanks the staff of BEPCII and the IHEP computing center for their strong support. This work is supported in part by National Key R\&D Program of China under Contracts Nos. 2020YFA0406300, 2020YFA0406400; National Natural Science Foundation of China (NSFC) under Contracts Nos. 11635010, 11735014, 11835012, 11935015, 11935016, 11935018, 11961141012, 12022510, 12025502, 12035009, 12035013, 12061131003, 12192260, 12192261, 12192262, 12192263, 12192264, 12192265, 12221005, 12225509, 12235017, 12150004, 12475197; Program of Science and Technology Development Plan of Jilin Province of China under Contract No. 20210508047RQ and 20230101021JC; the Chinese Academy of Sciences (CAS) Large-Scale Scientific Facility Program; the CAS Center for Excellence in Particle Physics (CCEPP); Joint Large-Scale Scientific Facility Funds of the NSFC and CAS under Contract No. U1832207; CAS Key Research Program of Frontier Sciences under Contracts Nos. QYZDJ-SSW-SLH003, QYZDJ-SSW-SLH040; 100 Talents Program of CAS; The Institute of Nuclear and Particle Physics (INPAC) and Shanghai Key Laboratory for Particle Physics and Cosmology; European Union's Horizon 2020 research and innovation programme under Marie Sklodowska-Curie grant agreement under Contract No. 894790; German Research Foundation DFG under Contracts Nos. 455635585, Collaborative Research Center CRC 1044, FOR5327, GRK 2149; Istituto Nazionale di Fisica Nucleare, Italy; Ministry of Development of Turkey under Contract No. DPT2006K-120470; National Research Foundation of Korea under Contract No. NRF-2022R1A2C1092335; National Science and Technology fund of Mongolia; National Science Research and Innovation Fund (NSRF) via the Program Management Unit for Human Resources \& Institutional Development, Research and Innovation of Thailand under Contract No. B16F640076; Polish National Science Centre under Contract No. 2019/35/O/ST2/02907; The Swedish Research Council; U. S. Department of Energy under Contract No. DE-FG02-05ER41374.

\end{document}